\documentclass[aps,prl,twocolumn,groupedaddress,showpacs,nofootinbib,longbibliography]{revtex4-1}
\usepackage{graphicx}
\usepackage{dcolumn}
\usepackage{bm}
\usepackage{hyperref}
\usepackage{upgreek}
\usepackage{color}
\usepackage{soul}
\usepackage{epstopdf}
\usepackage{units}
\usepackage{longtable}
\usepackage{floatrow}
\usepackage{mhchem}

\begin{document}
\title{Coexistence of surface and bulk ferromagnetism mimics skyrmion Hall effect in a topological insulator}

\author{K. M. Fijalkowski}
\affiliation{Faculty for Physics and Astronomy (EP3),
Universit\"at W\"urzburg, Am Hubland, D-97074, W\"urzburg, Germany}
\affiliation{Institute for Topological Insulators, Am Hubland, D-97074, W\"urzburg, Germany}
\author{M. Hartl}
\affiliation{Faculty for Physics and Astronomy (EP3),
Universit\"at W\"urzburg, Am Hubland, D-97074, W\"urzburg, Germany}
\affiliation{Institute for Topological Insulators, Am Hubland, D-97074, W\"urzburg, Germany}
\author{M. Winnerlein}
\affiliation{Faculty for Physics and Astronomy (EP3),
Universit\"at W\"urzburg, Am Hubland, D-97074, W\"urzburg, Germany}
\affiliation{Institute for Topological Insulators, Am Hubland, D-97074, W\"urzburg, Germany}
\author{P. Mandal}
\affiliation{Faculty for Physics and Astronomy (EP3),
Universit\"at W\"urzburg, Am Hubland, D-97074, W\"urzburg, Germany}
\affiliation{Institute for Topological Insulators, Am Hubland, D-97074, W\"urzburg, Germany}
\author{S. Schreyeck}
\affiliation{Faculty for Physics and Astronomy (EP3),
Universit\"at W\"urzburg, Am Hubland, D-97074, W\"urzburg, Germany}
\affiliation{Institute for Topological Insulators, Am Hubland, D-97074, W\"urzburg, Germany}
\author{K. Brunner}
\affiliation{Faculty for Physics and Astronomy (EP3),
Universit\"at W\"urzburg, Am Hubland, D-97074, W\"urzburg, Germany}
\affiliation{Institute for Topological Insulators, Am Hubland, D-97074, W\"urzburg, Germany}
\author{C. Gould}
\affiliation{Faculty for Physics and Astronomy (EP3),
Universit\"at W\"urzburg, Am Hubland, D-97074, W\"urzburg, Germany}
\affiliation{Institute for Topological Insulators, Am Hubland, D-97074, W\"urzburg, Germany}
\author{L. W. Molenkamp}
\affiliation{Faculty for Physics and Astronomy (EP3),
Universit\"at W\"urzburg, Am Hubland, D-97074, W\"urzburg, Germany}
\affiliation{Institute for Topological Insulators, Am Hubland, D-97074, W\"urzburg, Germany}

\date{\today}

\begin{abstract}
Here we report the investigation of the anomalous Hall effect in the magnetically doped topological insulator (V,Bi,Sb)$_2$Te$_3$. We find it contains two contributions of opposite sign. Both components are found to depend differently on carrier density, leading to a sign inversion of the total anomalous Hall effect as a function of applied gate voltage. The two contributions are found to have different magnetization reversal fields, which in combination with a temperature dependent study points towards the coexistence of two ferromagnetic orders in the system. Moreover, we find that the sign of total anomalous Hall response of the system depends on the thickness and magnetic doping density of the magnetic layer. The thickness dependence suggests that the two ferromagnetic components originate from the surface and bulk of the magnetic topological insulator film. We believe that our observations provide  insight on the magnetic behavior, and thus will contribute to an eventual understanding of the origin of magnetism in this material class. In addition, our data bears a striking resemblance to anomalous Hall signals often associated with skyrmion contributions. Our analysis provides a straightforward explanation for both the magnetic field dependence of the Hall signal and the observed change in sign without needing to invoke skyrmions, and thus suggest that caution is needed when making claims of effects from skyrmion phases.
\end{abstract}
\maketitle

The anomalous Hall effect (AHE), despite being first reported over a hundred years ago, \cite{Hall1881} remains of significant modern research interest as its investigation in novel magnetic materials continues to yield rich physics. Recent examples of this are the quantum anomalous Hall effect \cite{Liu2008,Yu2010,Qiao2010,Nomura2011,Chang2013,Checkelsky2014,Kou2014,Chang2015a,Bestwick2015,Grauer2015,Grauer2017}, which offers both potential for metrological applications \cite{Goetz2018,Fox2018} and for the academic study of axion electrodynamics \cite{Wilczek,Nomura2011,Grauer2017,Mogi2017}, as well as reports on alleged contribution to the Hall effect\cite{Yasuda2016,Vistoli2019,He2018,Wang2018,Liu2017,Matsuno2016,Ludbrook2017,Ohuchi2018} associated with skyrmion magnetic textures.

Our current investigation pertains to both these observations, as it aims to both deepen our understanding of the nature of the magnetic state in materials where the quantum anomalous Hall effect has been reported, and shows that signals currently identified as a contribution from skyrmions are potentially being misinterpreted since, as we show here, the same signal can be more easily interpreted as simply resulting from the interplay between two ferromagnetic contributions in a material which does not host skyrmions.

Our studies are carried out on layers of the magnetic topological insulator V-doped (Bi,Sb)$_2$Te$_3$. Most works superficially describe its magnetic state as simply ferromagnetic, but some show indications of superparamagnetism \cite{Grauer2015}, magnetically fluid behavior, or even claims of adiabatic cooling \cite{Bestwick2015}. Moreover, a clear explanation of the origin of the strong perpendicular-to-plane magnetic anisotropy, and of the relative difference in the strength of this anisotropy for V vs Cr doping \cite{Chang2015a} is still lacking.

In this paper, we aim to shed more light on the nature of the magnetic state through a study of the AHE \cite{Smit1955,Smit1958,Berger1964,Jungwirth2002,Onoda2002,Nagaosa2010} in these materials. We find that it shows indications of a rich magnetic phase resulting from the interplay between two individual magnetic components.     

All investigations are performed on samples with a standard Hall bar geometry prepared by optical lithography, and are equipped with an electrostatic gate that allows tuning of the Fermi level in the layers. The topological insulator heterostructures  are (Bi,Sb)$_2$Te$_3$ based, with the magnetic layers doped by V. They are grown using molecular beam epitaxy (MBE) on insulating, hydrogen passivated Si (111) substrates, and are covered in-situ with 10 nm of Te to protect the surface from aging \cite{Winnerlein2016,Grauer2017}. The Bi/Sb ratio is 0.79/0.21 for all samples. More details on the growth and fabrication can be found in the Supplementary material\cite{Suppl}. Measurements are done in basic \ce{^{4}_{}He} magnetocryostats at temperatures ranging from 2.4 K to 13 K and unless otherwise specified, at the applied gate voltage yielding the highest value of $\rho_{xx}$.

\begin{figure}[]
\includegraphics[width=\columnwidth]{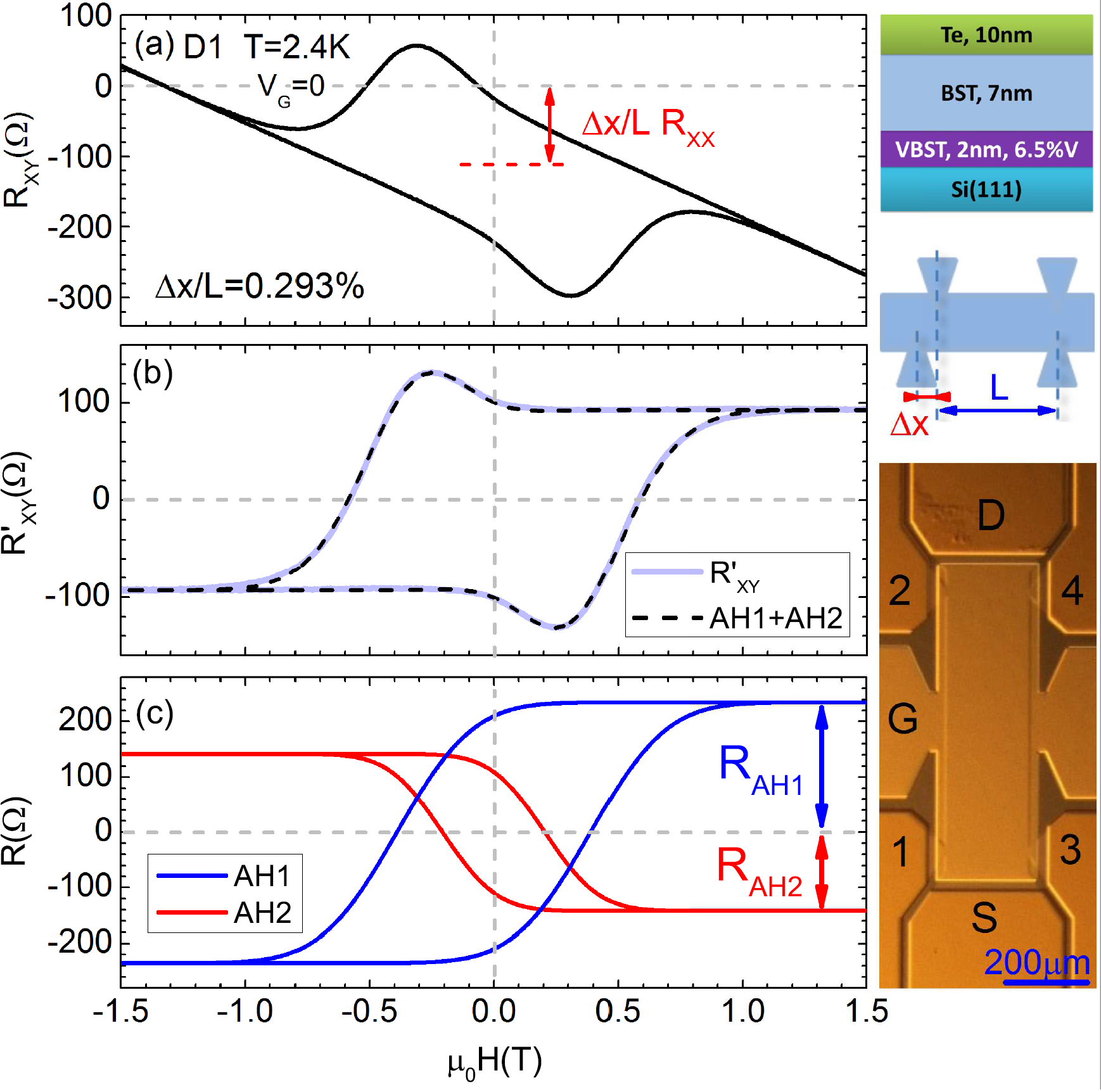}
\caption{
a) Measurement of the Hall resistance as a function of H field for device D1 (layer schematic on the right). The red arrow indicates the offset resulting from mixing in of the longitudinal resistance (see text for details). b) Solid line: Same data after removing offset and classical Hall contributions as described in the text. Dashed line: result of the two component model used to describe the data. c) each of the two anomalous Hall contributions extracted from fitting the data. To the right of the figure are a picture of one of the devices, and a schematic describing how we model imperfections in the device (see text).
}%
\label{fig:Fig1}%
\end{figure}
\begin{figure}[]
\includegraphics[width=\columnwidth]{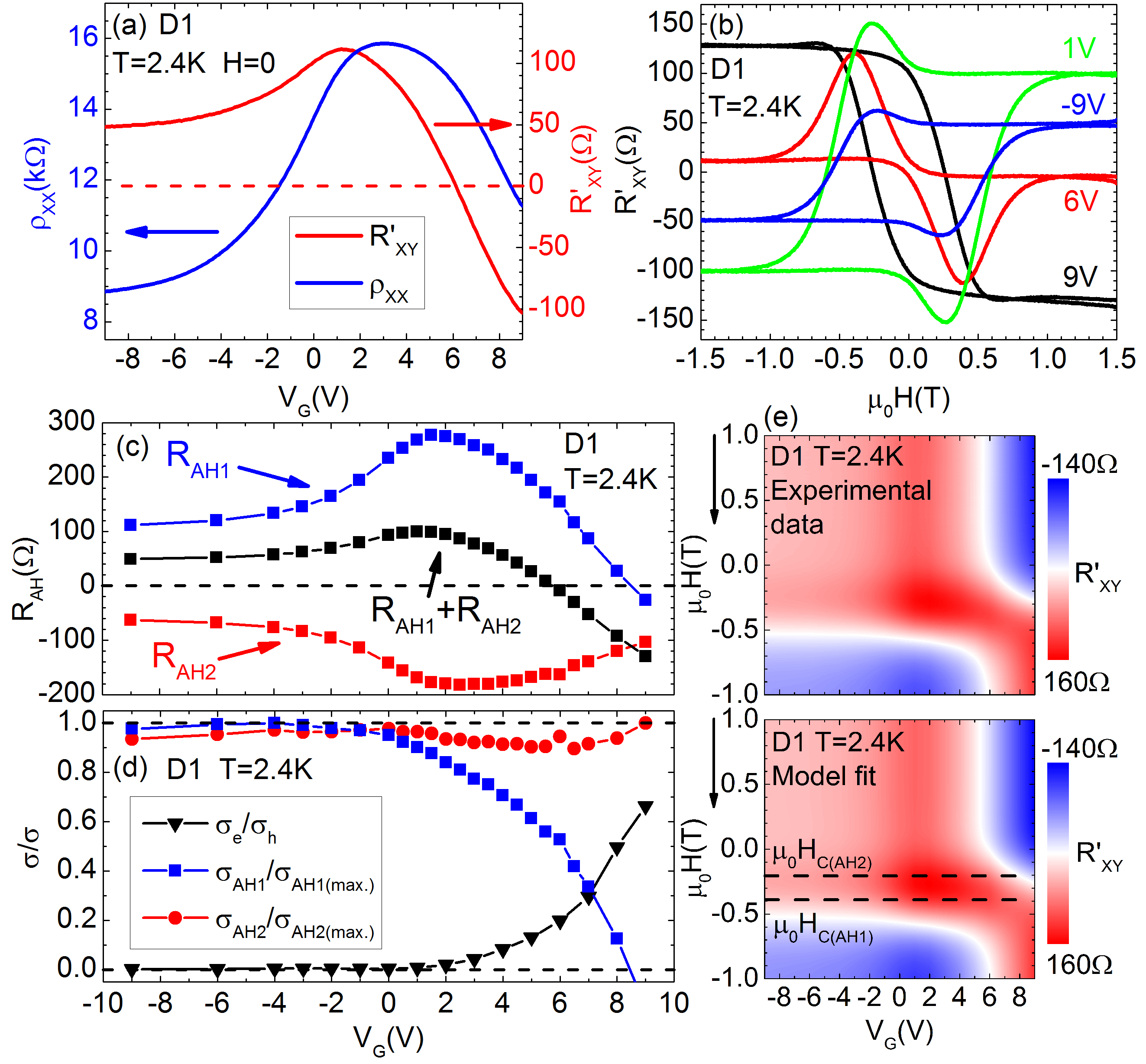}
\caption{
a) Zero-field $\rho_{xx}$ and R'$_{xy}$ of our device as a function of gate voltage. b) Magnetic field dependence of the offset corrected Hall resistance (R'$_{xy}$) as a function of magnetic field for various gate voltages.  c) Strength of each of the contributions extracted from the data. d) The two normalized anomalous Hall conductivity components as a function of gate voltage compared to the ratio between electron and hole Drude conductivity extracted from the high field data (see Supplementary material\cite{Suppl}). e) Color plot confirming the agreement between the experimental data (top) and the description provided by our model (bottom).
}%
\label{fig:Fig2}%
\end{figure}

Our main investigations are focused on device D1, which was selected because it most clearly illustrates the rich nature of the AHE behavior in these materials, while at the same time showing a total anomalous Hall signal which is very reminiscent of the one associated with the topological skyrmion term. It comprises of a 9 nm thick TI layer, the bottom 2 nm of which contain 6.5$\%$ V doping. Fig.1a shows the AHE response of this device at a temperature of 2.4 K and with zero gate voltage applied. Its shape is significantly more featureful than expected from a simple ferromagnet, and as we will demonstrate, is the result of the co-existence of two ferromagnetic contributions in the sample.

Despite being produced by high precision lithography, in any sample where the longitudinal resistance is larger than the Hall resistance, a small ad-mixing of R$_{xx}$ into the Hall signal is inevitable. This explains why the hysteresis loop in Fig.1a is not vertically symmetric around the origin. To carefully analyze the Hall signal, we first remove this small additional contribution by treating ad-mixing (which always results from the fact that the Hall voltage is not being picked up strictly orthogonal to the current flow) as an effective misalignment of the Hall leads, as illustrated (strongly exaggerated) in the sketch of Fig. 1. The pure Hall signal can then be obtained by subtracting away the correct fraction of the longitudinal voltage (conceptually R$_{xx}$*$\Delta$x/L), where in our case $\Delta$x/L is always smaller than $0.005$. The resulting total Hall data contains not only the AHE, but also the classical Hall Effect, which can be removed by subtracting the appropriate linear slope (a positive slope corresponds to n-type transport). The result of removing both the ad-mixing and the classical Hall Effect leaves us with the pure AHE contribution, which we label as R'$_{xy}$ to distinguish from the measured values, and which is presented as the pale solid line in Fig.1b.

These corrections obviously change the quantitative details of the curve, but do not modify its overall qualitative appearance, which remains quite different from that expected from a simple ferromagnet. In fact, the curve can be decomposed into two contributions, each having the appearance of a simple AHE signal and having anomalous Hall coefficients of opposite signs. This decomposition is displayed in Fig.1c, where the blue and red lines show the two individual contributions, one positive (AH1) and the other negative (AH2). The sum of these two is shown as the dashed line overlaid to the data in Fig.1b. We adopt the convention that the negative anomalous Hall contribution is the one which has the same sign as the quantum anomalous Hall effect displayed by our samples when cooled to below 100 mK (see Supplementary material\cite{Suppl}). We therefore refer to the contribution described by the red line in Fig.1c as negative.

Because this device is gated, we can now study the evolution of these two components as we change the carrier concentration in the layer. In fig.2a we plot the H=0, $\rho_{xx}$ and R'$_{xy}$ value for the device as a function of gate voltage, showing that the device can be gated through its minimum conductivity point. In fig.2b, we show R'$_{xy}$ (the measured R$_{xy}$ curves are given in fig.S6 of the Supplementary material\cite{Suppl}) as a function of magnetic field for different gate voltages. Interestingly, not only does the shape of the curve change, but even the overall sign of the AHE reverses as we go from large negative to large positive gate voltages. In all cases, the data can be modeled by decomposing it into two contributions as in Fig.1c. The agreement between the resulting fits and the data is illustrated in fig.2e, where the upper panel is a color scale of the experimental R'$_{xy}$ as a function of magnetic field and gate voltage, and the lower panel is the same plot, as generated by the model. 

The strength of the two contributions determined by this fit is presented as the red and blue points in Fig.2c, with the black points giving the sum of the two, and reproducing the fact that the sign of the overall AHE changes at around 6V of applied gate voltage. The coercive fields of the positive and negative contributions, defined as the magnetic field where the contribution crosses zero, are 390 mT and 206 mT, respectively, and have no dependence on gate voltage. Using standard matrix inversion, these results can be converted to anomalous Hall conductivity contributions, yielding the red and blue points of Fig.2d. We compare these to the black points in the figure, which we get from fits of the high field classical Hall signal (see Supplementary material\cite{Suppl} for details). It is noteworthy that while the negative anomalous Hall contribution is relatively independent of gate voltage, the positive one decreases quickly as electron conductivity begins to turn on.

\begin{figure}[]
\includegraphics[width=\columnwidth]{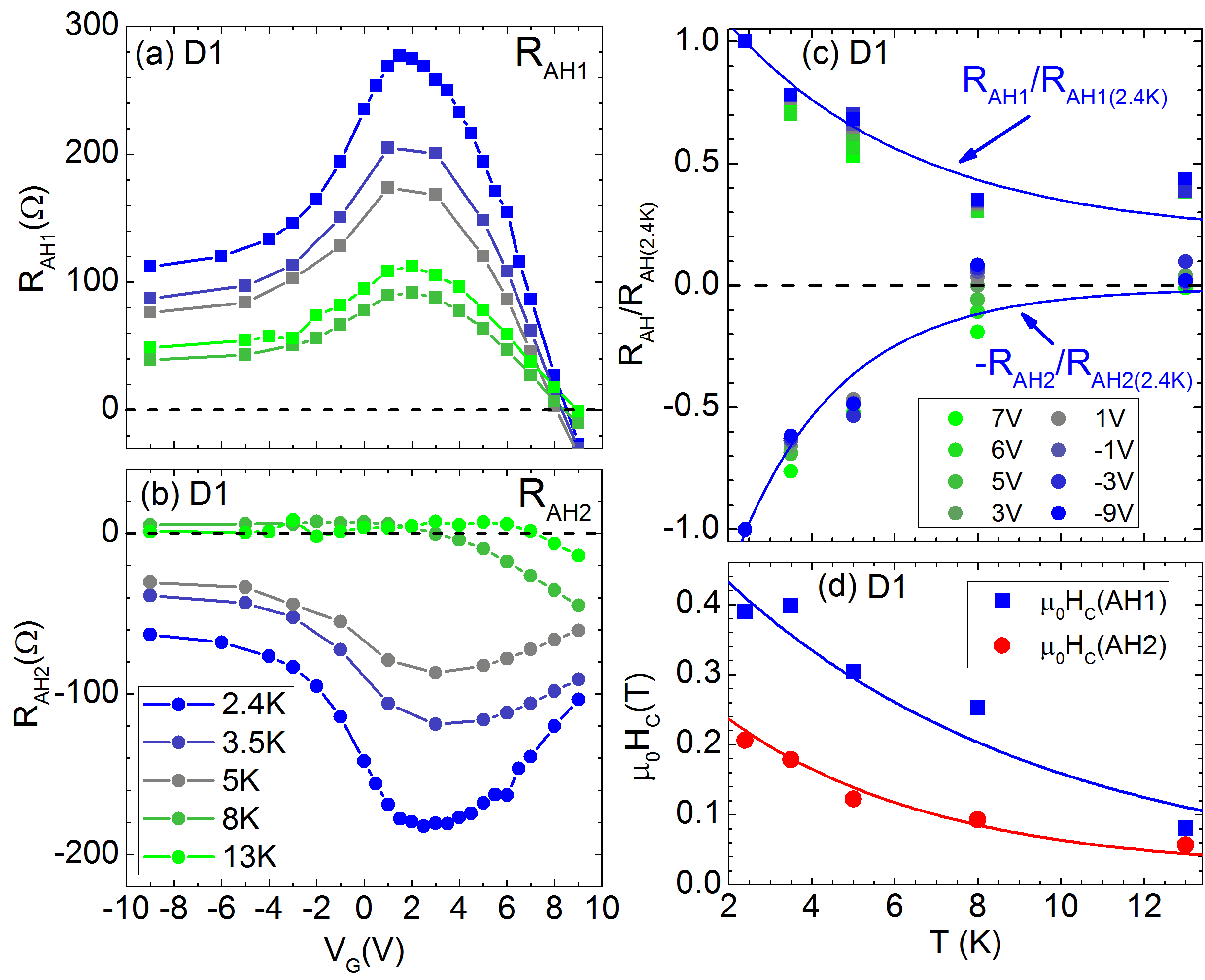}
\caption{a)/b) gate voltage dependence of the positive/negative contribution for measurements at various temperatures. c) Temperature dependence of each of the (normalized) components, and d) temperature dependence of the coercive field associated with each of the component. (The lines are guides to the eye.)
}
\label{fig:Fig3}%
\end{figure}

\begin{figure}[]
\includegraphics[width=\columnwidth]{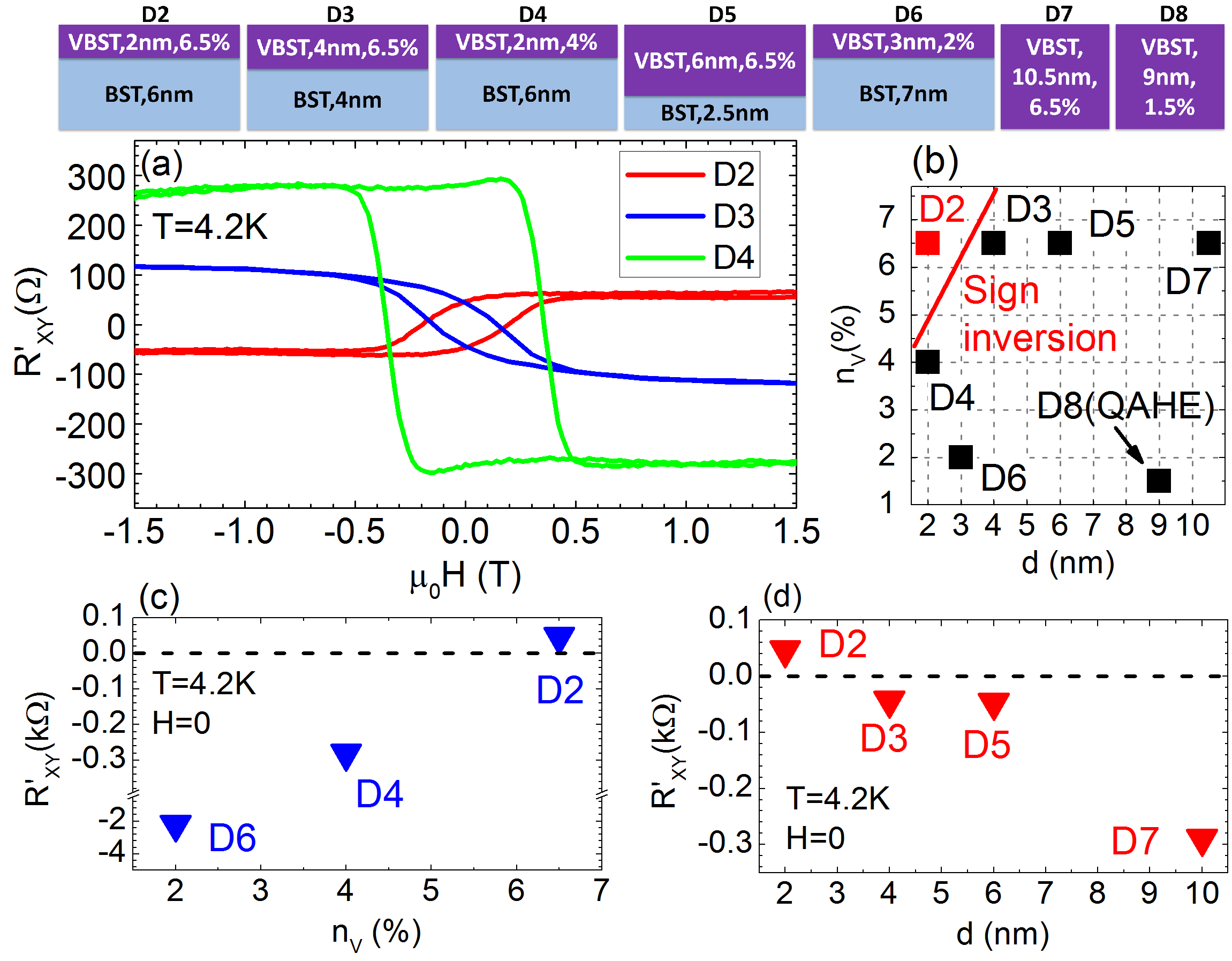}
\caption{
a) Magnetic field dependence of the offset corrected Hall resistance (R'$_{xy}$) for devices D2-D4. b) Magnetic doping / thickness phase space with position of the devices D2-D8, including the position where the quantum anomalous Hall effect was observed in this material system (device D8, see Supplementary material\cite{Suppl}). c)/d) Offset corrected zero field value of the Hall resistance as a function of magnetic doping level (c) and magnetic layer thickness (d) for the positively magnetized state. A schematic layer layout for devices D2-D8 is presented at the top of the figure.
}
\label{fig:Fig4}%
\end{figure}

Further evidence of the relative independence of these two contributions is provided by a temperature dependence study. The same analysis as in fig.2 is now conducted on the sample at temperatures ranging from 2.4 K to 13 K, and the results are summarized in Fig.3. In panel a and b, we plot the amplitude of each contribution as a function of gate voltage, and for several temperatures. From this data, we plot in Fig.3c the normalized amplitude vs T for a few characteristic gate voltage values, and in Fig.3d, the coercive field of each component at a few temperatures, again showing the clearly different behavior of the two components.

Device D1 was ideal for this detailed study because it has both negative and positive contributions to its AHE which are similar enough in amplitude yet different enough in coercive fields that both can be reliably extracted for the analysis, and also because it clearly highlights the similarities between our results and those attributed to skyrmions. A study of more and varied samples allows us additional insights, and indeed suggests that in our samples, the two ferromagnetic states can be associated with surface and bulk contributions. To verify this, we have examined a large set of samples where the thickness and composition of the magnetic layer is varied. A sampling of these results is presented in Fig.4 (with the results for still more samples provided in the Supplementary material\cite{Suppl}).

As seen in Fig.4a, while some devices (D4) show both contributions as did D1, other samples have anomalous Hall signal which are dominated by a single contribution. Even there however, the situation is non-trivial, as the AHE can be either positive (D2) or negative (D3) depending on the properties of the individual layer. As evident from the summary of the results given in Fig 4b-d, it appears that the positive contribution is driven from very thin and highly magnetically doped layers.

For a constant high magnetic doping level, positive total AHE is observed only when the magnetically doped film thickness is sufficiently small. When thickness is manipulated under a constant magnetic doping concentration (see fig.4d), the amount of magnetic dopants on the film surface remains unaffected (the two main surfaces are only shifted with respect to each other), whereas the amount of magnetic dopants in the bulk grows as the two surfaces shift away from each other. Based on that we suggest that the positive contribution originates on the surface of the magnetically doped layer, and the negative one in the bulk. This assignment is also supported by the fact that the positive contribution is much more sensitive to gate voltage than the negative one. Moreover, the different coercive fields and temperature dependence of each component indicate separate ferromagnetic origin, pointing towards the coexistence of surface and bulk ferromagnetism. It is not surprising that different magnetic and charge environments (especially with regard to a topological surface state) for the magnetic dopants on the surface and in the bulk would result in different magnetic interactions leading to the observed coexistence of two magnetic orders. The exact microscopic origin of the distinct surface and bulk magnetism remains an open question, and one that, when answered, may provide deep insight as to the underlying nature of the, still badly understood, magnetism in this material. 

We note that all published transport results comprising similar phenomenology\cite{Yasuda2016,Vistoli2019,He2018,Wang2018,Liu2017,Matsuno2016,Ludbrook2017,Ohuchi2018} could be well explained using our model introducing two coexisting ferromagnetic orders. Moreover our explanation successfully describes every electrical transport feature, including dependence on magnetic field, gate voltage, temperature and thin film parameters such as doping level and film thickness, without the need to speculate about new topological physics. While our analysis can't explicitly rule out a skyrmion origin to the observed transport features in some cases, it certainly provides the simplest explanation yet, and implies that extreme caution should be used when making claims of effects from skyrmion phases.

{\it Conclusions.} 
We have investigated the magneto-transport properties of magnetically doped (V,Bi,Sb)$_2$Te$_3$ topological insulator structures. Rich anomalous Hall response exhibiting sign reversal as a function of magnetic layer thickness, magnetic doping density and applied gate voltage was observed. All transport features can be well explained by the coexistence of two ferromagnetic orders in the system. Moreover, the analysis suggests that one of the ferromagnetic orders originates on the surface while the other in the bulk of the magnetic topological insulator film. Our work provides yet another valuable piece to the puzzle of the origin of magnetism in this material system, and we believe will contribute to a final understanding of physics underlying it. Moreover, the characteristic shape of the anomalous Hall signal in the case where the two contributions are comparable in amplitude, as well as the observed change of overall sign of the signal as a function of parameters such as gate voltage and temperature comprise an identical phenomenology to that commonly attributed to skyrmion states. Our findings therefore suggest the necessity for reexamination of these earlier claims.

\begin{acknowledgments}
We gratefully acknowledge the financial support of the EU ERC-AG Program (project 3-TOP), the EU ERC-StG Program (project TOPOLECTRICS), the DFG (SFB 1170, 258499086), and the Leibniz Program. We thank Alexandra Sutherland for the help with the samples fabrication.
\end{acknowledgments}

\bibliography{Signinv6}

\newpage

\setcounter{figure}{0}
\renewcommand{\thetable}{S\arabic{table}}  
\renewcommand{\thefigure}{S\arabic{figure}} 

\huge
\centerline{\textbf{Supplementary material}}
\normalsize

\section{1. Sample fabrication.}
All investigated (V,Bi,Sb)$_2$Te$_3$ layers were grown using MBE on insulating hydrogen passivated Si(111) substrates and capped in-situ with 10 nm of Te (except devices D7 and D14) as a protection from ambient conditions and the chemicals used in the lithographic process. The Te cap is insulating at measurement temperature. The MBE system is equipped with thermal effusion cells and the substrate temperature during the growth was monitored by a thermocouple on the backside of the substrate heater in the same distance as the PBN plate to which the substrate is clamped. \cite{Winnerlein2016} The temperature read by the thermocouple during the growth of all of the layers was kept constant at 190$^{\circ}$C. The Bi/Sb ratio was determined by X-ray diffraction (XRD) measurements of the lateral lattice constant a.\cite{Winnerlein2016} It was determined to be 0.21/0.79 in a 9 nm thick film with uniform V doping of $1.5~\%$ of all atoms (device D8 exhibiting the quantum anomalous Hall effect, see section 6). Selective magnetic doping was performed by the time controlled opening and closing of the vanadium cell shutter during the MBE growth, and the magnetic doping concentration was controlled by the vanadium cell temperature. The exact value of the magnetic doping content was determined from the growth rate of pure vanadium and reference samples. 

The samples are patterned using standard optical lithography. In a first lithographic step, two six-terminal Hall bar mesas are defined using Ar ion beam etching (see microscope images in the Fig.S1) with the use of either a polymer photo-resist mask, or a hard titanium mask. In a second step the top gate oxide consisting of 20 nm of AlOx, and 1 nm of HfOx is deposited using atomic layer deposition (ALD), and then immediately covered with the photo-resist, developed, and transferred into a high vacuum e-beam evaporation metallization chamber, where a gate electrode stack of 2 nm of Ti and 100 nm of Au is deposited. In a third step the ohmic AuGe contact leads are defined. For that, the gate oxide is removed from the contact area using diluted HF. The sample is then transferred into a high vacuum cluster, where without breaking the high vacuum conditions the Te cap is removed from the contact area with Ar ions, and 50 nm of AuGe alloy followed by 50 nm of Au is deposited by electron beam evaporation. Every sample contains two Hall bar devices with widths 10 $\mu$m and 200 $\mu$m and aspect ratio 3/1. After the lithographic process each sample is glued to a chip carrier, and the wedge bonding method is used to connect the leads on the device with the leads on the chip carrier through Au wires.

\section{2. Microscope images of one of the samples.}
Fig.S1 shows various Nomarski microscope images of one of the samples.

\section{3. Hall magneto-resistance data for devices D2-D14.} Fig.S2 shows the Hall magneto-resistance data for devices D2-D14 collected at 4.2 K for the gate voltage providing the highest value of $\rho_{xx}$ in each of the device. Table S1 displays layer stack details for every device.

\section{4. Longitudinal magneto-resistivity data for devices D2-D14.} Fig.S3 shows the longitudinal magneto-resistivity data for devices D2-D14 collected at 4.2 K and corresponding to the Hall data in Fig.S2. 

\section{5. Gate voltage dependence for devices D2-D6,D8-D14.} Fig.S4 presents the gate voltage sweeps collected at 4.2 K for devices D2-D6,D8-D14. Measurements were performed at zero external magnetic field (except for device D6 where the gate voltage sweep was performed at the applied external magnetic field of $\mu_{0}$H=-0.5 T). Every device was magnetically prepared in order to align the magnetic moments in the material before the gate voltage sweep. The indices M+(M-) indicate that the external magnetic field was set to $\mu_{0}$H=2 T($\mu_{0}$H=-2 T) before returning it to zero and performing the experiment. None of those devices exhibits an anomalous Hall sign tunability with the gate voltage, likely because they cannot be tuned to a sufficiently n-type transport regime.

\section{6. Quantum anomalous Hall effect in device D8} Fig.S5 demonstrates the quantum anomalous Hall state observed in one of the investigated devices (D8). A hysteresis loop measurement performed at a temperature 25 mK demonstrates switching between the two edge state chiralities (by reversing the magnetization direction) and shows precisely quantized value of $\rho_{xy}=\pm$h/e$^2$ and $\rho_{xx}=0$.

\begin{table}
\begin{tabular}{|| c || c | c | c | c | c | c | c | c |}
\hline
 & \textbf{d$_{1}$} & \textbf{n$_{1}$} & \textbf{d$_{2}$} & \textbf{n$_{2}$} & \textbf{d$_{3}$} & \textbf{n$_{3}$} & x/y \\ \hline
\textbf{D1} 	& 2 	& 6.5 & 7 	& 0 	& - 	& -  	&	200x600 $\mu$m$^2$ 	\\ \hline
\textbf{D2} 	& 6 	& 0 	& 2 	& 6.5 & - 	& - 	&	200x600 $\mu$m$^2$ 	\\ \hline
\textbf{D3} 	& 3.5 & 0 	& 4 	& 6.5 & - 	& - 	&	10x30 $\mu$m$^2$ 		\\ \hline
\textbf{D4} 	& 6 	& 0 	& 2 	& 4 	& - 	& - 	&	10x30 $\mu$m$^2$ 		\\ \hline
\textbf{D5} 	& 2.5 & 0 	& 6 	& 6.5 & - 	& - 	&	10x30 $\mu$m$^2$ 		\\ \hline
\textbf{D6} 	& 7 	& 0 	& 3 	& 2 	& - 	& - 	&	10x30 $\mu$m$^2$ 		\\ \hline
\textbf{D7} 	& 10.5& 6.5 & - 	& - 	& - 	& - 	& 200x600 $\mu$m$^2$ 	\\ \hline
\textbf{D8} 	& 9 	& 1.5 & - 	& - 	& - 	& - 	& 200x600 $\mu$m$^2$ 	\\ \hline
\textbf{D9} 	& 3.5 & 0 	& 2 	& 6.5 & 3.5 & 0 	&	10x30 $\mu$m$^2$ 		\\ \hline
\textbf{D10} 	& 2 	& 6.5 & 7 	& 0 	& - 	& - 	&	10x30 $\mu$m$^2$ 		\\ \hline
\textbf{D11} 	& 2 	& 6.5 & 4.5 & 0 	& 2 	& 6.5 & 10x30 $\mu$m$^2$ 		\\ \hline
\textbf{D12} 	& 2 	& 6.5 & 6.5 & 0 	& 2 	& 4		&	200x600 $\mu$m$^2$ 	\\ \hline
\textbf{D13} 	& 2 	& 6.5 & 4.5 & 0 	& 2 	& 4		&	200x600 $\mu$m$^2$ 	\\ \hline
\textbf{D14} 	& 10 	& 2.7 & - 	& - 	& - 	& - 	&	10x30 $\mu$m$^2$ 		\\ \hline
\end{tabular}
\caption{Table presenting thicknesses and magnetic doping contents for individual layers in the heterostructure stack for all devices presented in the paper and supplementary material (devices D1-D14). Parameters d$_{1,2,3}$ and n$_{1,2,3}$ denote thickness and percentage of V atoms (as a fraction of all atoms) in each of the layers, and an entry of "-" indicates that the layer is not present in the hetero-structure. The $x/y$ column shows the Hall bar dimensions for each of the investigated device.}%
\label{tab:TabS1}%
\end{table}

\section{7. Magneto-resistance data for our representative device D1} 
Fig.S6a shows typical raw Hall magneto-resistance data collected at 2.4 K for gate voltages ranging from V$_G$=-9 V (yellow line) to V$_G$=9 V (blue line). The plotted gate voltage values are V$_G$=-9 V, V$_G$=-6 V, V$_G$=-4 V, and then in an increment of 1 V to V$_G$=9 V. For further analysis, the data has been corrected for ad-mixing as described in the main text, and plotted in Fig.S6b ($\rho_{xx}$ data plotted in Figs.S6g/h). The subtracted ordinary Hall effect background is plotted in Fig.S6c (blue points) and compared to the values calculated from the high field analysis for consistency (red points). The anomalous Hall hysteresis loops (after the voltage crosstalk correction and the Hall slope removal) are plotted for clarity in three separate figures in Fig.S6d/e/f. This data set was analyzed to obtain the anomalous Hall amplitudes plotted in Fig.2 of the main text.

\section{8. Details of the double anomalous Hall model.} 
To simulate both ferromagnetic hysteresis loops and obtain the parameters presented in Figs.1, 2 and 3 of the main text, functions of the form R$_{xy}$=R$_{AH}$erf(A (H-H$_C$)) were used, where R$_{AH}$ is the anomalous Hall amplitude, H$_C$ is the switching field of the ferromagnet, and A is the reversal width. For every temperature, parameters H$_C$ and A were fixed and only the anomalous Hall amplitudes were used as fitting parameters. Typical fitting results are demonstrated in Figs.S7a-e. Black solid lines show the R'$_{xy}$ data for a few gate voltage values, and olive solid lines show the respective fit curves composed of two ferromagnetic switching events. In Figs.S7f-j the two separate ferromagnetic contributions are plotted for each gate voltage. The red solid lines represent the negative contribution, and the blue lines show the positive contribution. 

\section{9. High external magnetic field data and multiple carrier analysis.} 
To obtain the data presented in Fig.2d of the main text (black points) we performed high magnetic field analysis for device D1 at a temperature of 2.4 K. Figs.S8a/b show respectively the typical $\rho_{xx}$ and R$_{xy}$ gate voltage sweeps collected at fixed values of the external magnetic field ranging from $\mu_{0}$H=14 T (blue line) to $\mu_{0}$H=1 T (yellow line) in intervals of 1 T.

To investigate the magnetic field dependence of R$_{xy}$ the gate voltage sweeps collected at different fixed magnetic field values were transposed to get the magnetic field dependence for different fixed gate voltage values. The same procedure was performed for the $\rho_{xx}$ gate sweeps. The R$_{xy}$ data was then corrected for the voltage crosstalk, with the resulting R'$_{xy}$ curves plotted in the Fig.S8c/d. The solid lines represent the experimental data ranging from gate voltage V$_G$=1 V (yellow color) to V$_G$=9 V (blue color) in Fig.S8c, and from gate voltage V$_G$=-9 V (yellow color) to V$_G$=0 V (blue color) in Fig.S8d. The fitted two-carrier model curves (dashed lines) are overlaid on the data in Figs.S8c/d. Fig.S8e shows the conductivities: Drude zero-field conductivities for electrons (green points, definition in the legend) and holes (red points, definition in the legend), the sum of both (dark yellow points), and the experimental zero-field value of $\sigma_{xx}$ (blue points). Fig.S8f shows the obtained sheet densities of electrons (blue points), holes (red points) and the total density (black points). The individual mobilities of electrons and holes are plotted in Fig.S8g, and ratios between the different parameters: electron/hole sheet density ratio (Fig.S8h), electron/hole mobility ratio (Fig.S8i) and electron/hole Drude zero-field conductivity ratio (Fig.S8j; the same data as the black points in Fig.2d of the main text).

\section{10. Temperature evolution of the magneto-resistance in the device D1.} 
Fig.S9 shows Hall magneto-resistance and longitudinal magneto-resistivity for a wide range of fixed gate voltages and temperatures ranging from 3.5 K to 13 K, from which the parameters presented in the Fig.3 were determined. Typical corrected R'$_{xy}$ hysteresis loops are presented in the Fig.S10 for better visibility.

\onecolumngrid
\noindent\rule[0.5ex]{\linewidth}{1pt}

\begin{figure}[h]%
\includegraphics[width=\columnwidth]{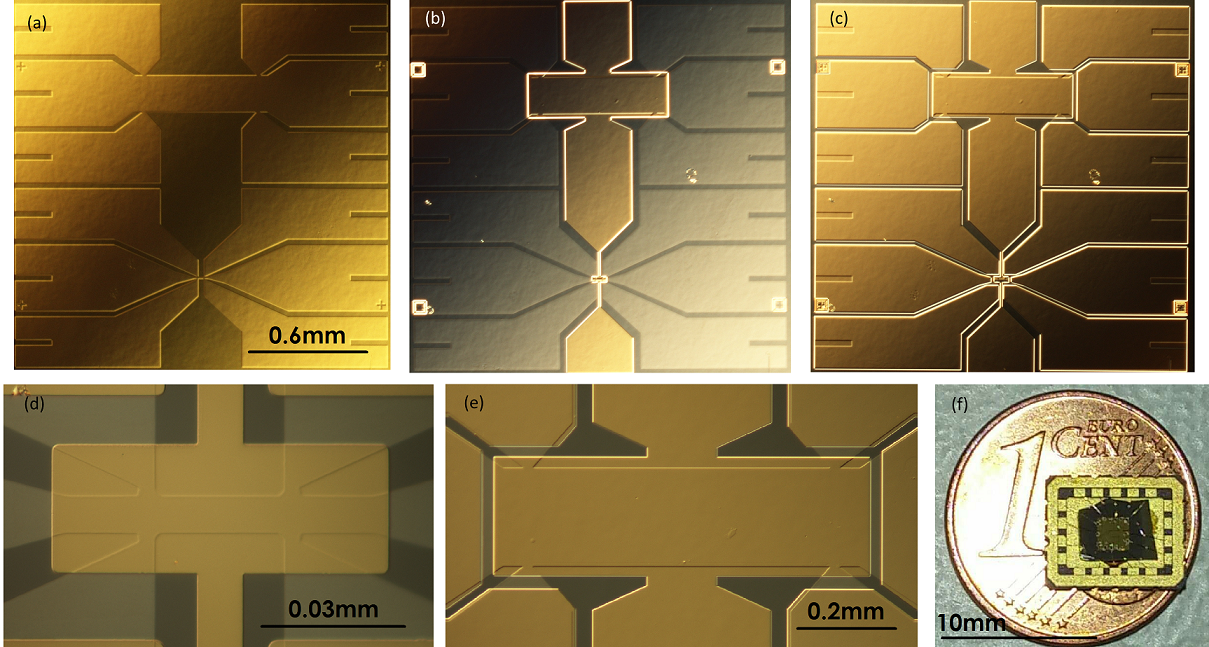}
\caption{(a) Differential contrast interference microscope images of one of the processed samples. Image after the mesa definition, (b) image after the gate electrode definition, (c) image of the finished sample after the contact leads definition (image magnification is the same in (a), (b), and (c)), (d) images of the finished small Hall bar device, (e) the big Hall bar device, and (f) the photo of the processed samples lying on top of the 1 Eurocent coin, consisting of the processed sample glued and electrically connected to the chip carrier.}%
\label{fig:Figs1}%
\end{figure}

\begin{figure}[h]%
\includegraphics[width=\columnwidth]{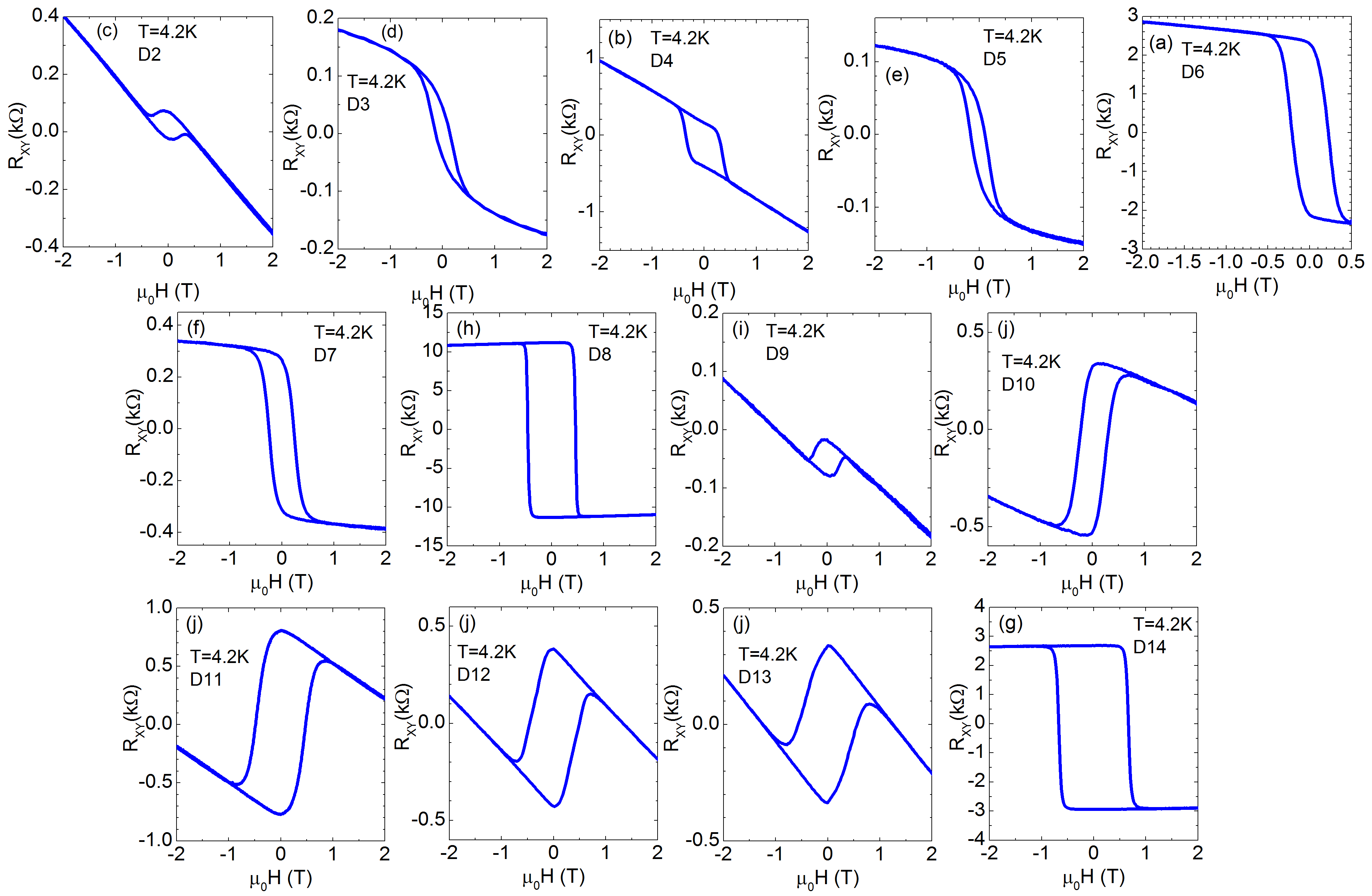}
\caption{
Raw Hall magneto-resistance from devices D2-D14 collected at 4.2 K for the gate voltage providing the highest $\rho_{xx}$ in each sample. 
}%
\label{fig:Figs2}%
\end{figure}

\begin{figure}[h]%
\includegraphics[width=\columnwidth]{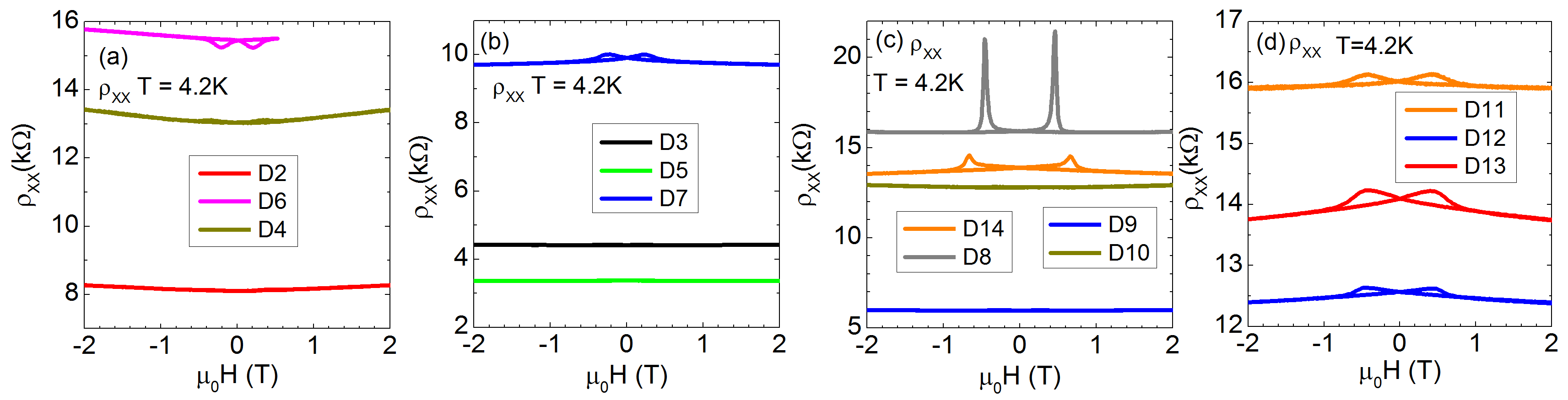}
\caption{
Raw longitudinal magneto-resistivity data from the devices D2-D14 collected at 4.2 K for the gate voltage providing the highest $\rho_{xx}$ in each sample. 
}%
\label{fig:Figs3}%
\end{figure}

\begin{figure}[h]%
\includegraphics[width=\columnwidth]{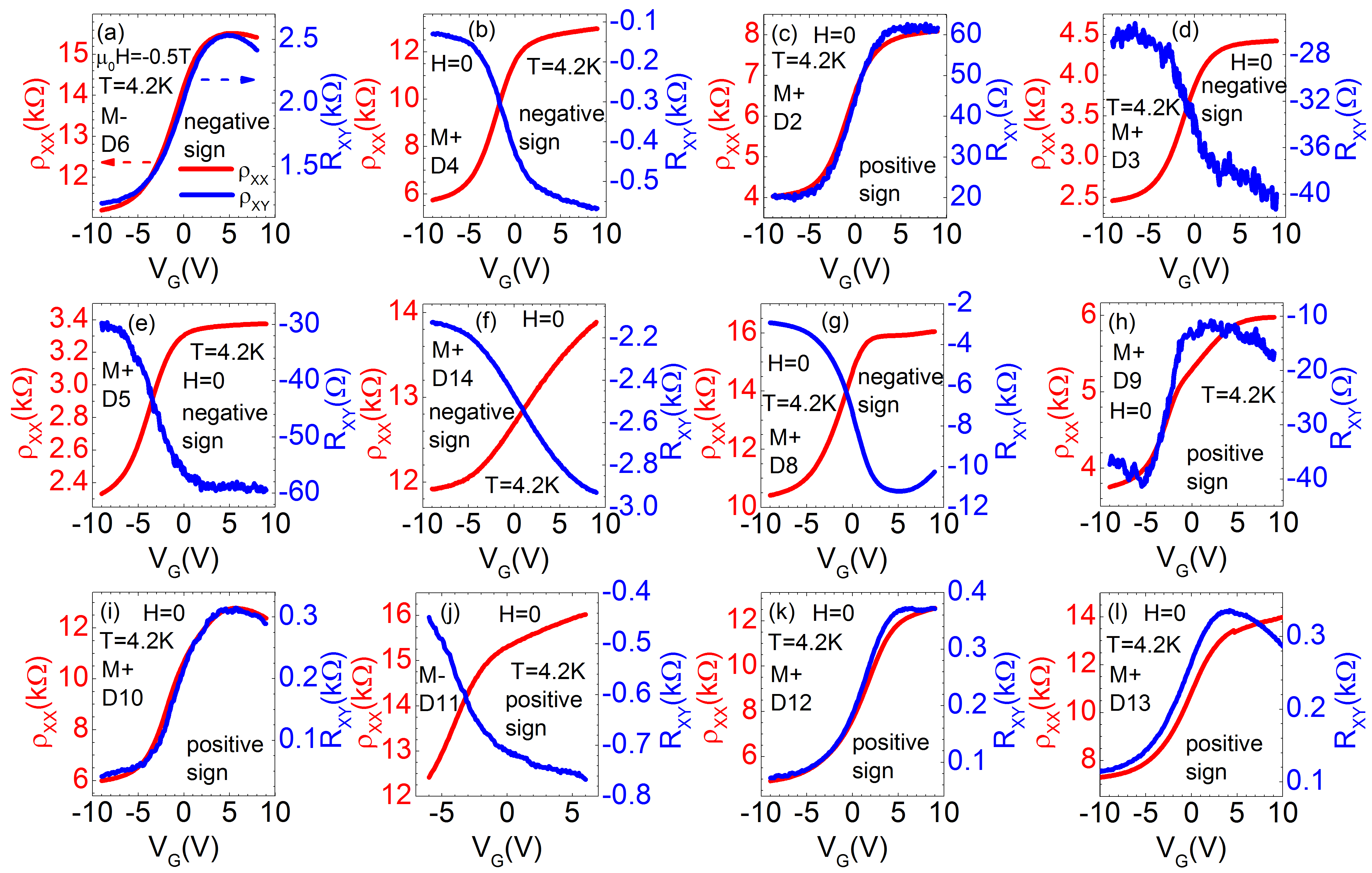}
\caption{
Raw experimental data showing gate voltage sweeps of the longitudinal and Hall resistivity of devices D2-D6,D8-D14 collected at 4.2 K and no external magnetic field H=0 (except for device D6 where the gate voltage sweep was performed with an external magnetic field of $\mu_{0}$H=-0.5 T). Every device was magnetically prepared in order to align the magnetic moments in the material before the gate voltage sweep. The indices M+(M-) indicate that the external magnetic field was set first to the value of $\mu_{0}$H=2 T($\mu_{0}$H=-2 T) before setting it back to 0 and performing the experiment. The red color represents the $\rho_{xx}$ data, the blue color represents the R$_{xy}$ data, the colors of the labels on the axes coincide with the colors of the curves.
}%
\label{fig:Figs4}%
\end{figure}

\begin{figure}[h]%
\includegraphics[width=\columnwidth]{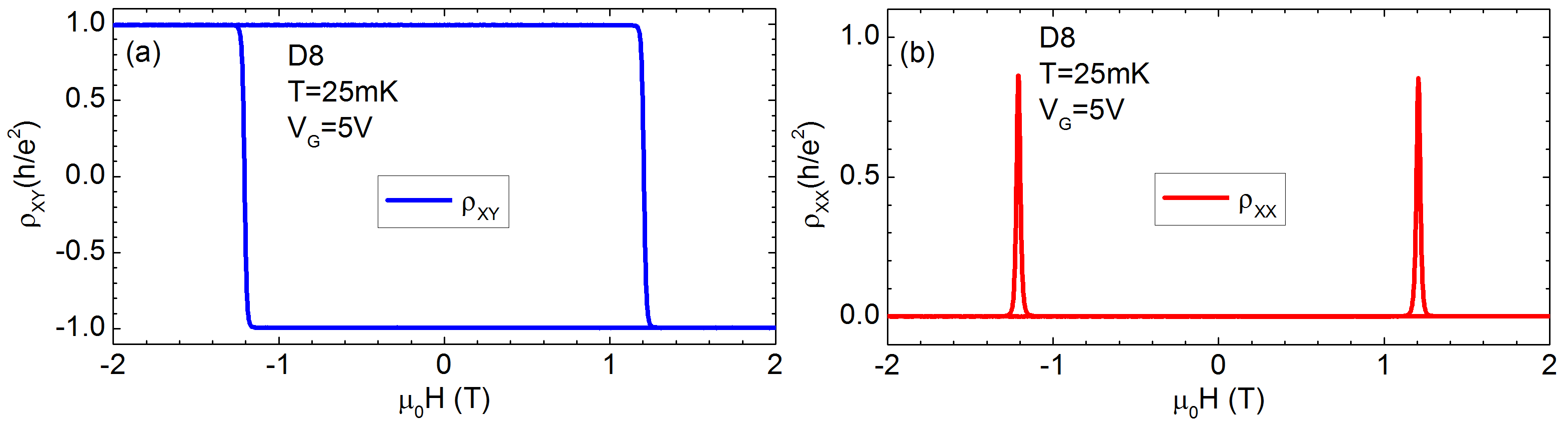}
\caption{
Quantum anomalous Hall effect in the device D8. (a) Hall and (b) longitudinal resistivity collected at 25 mK demonstrating a perfect quantum anomalous Hall state in device D8.
}%
\label{fig:Figs5}%
\end{figure}

\begin{figure}[h]%
\includegraphics[width=\columnwidth]{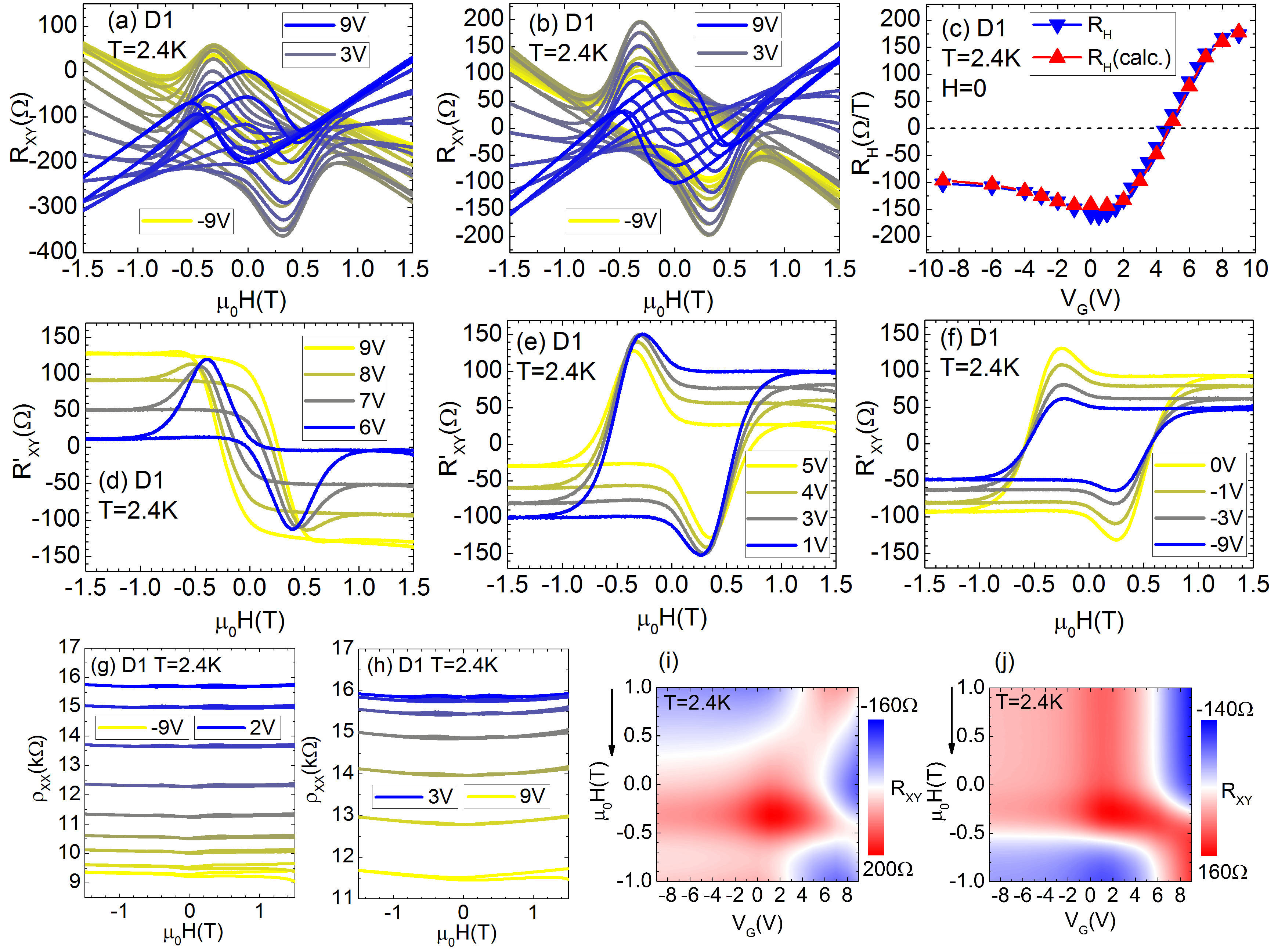}
\caption{
Data from device D1 collected at 2.4K. (a) Raw Hall magneto-resistance for a full range of fixed gate voltage values from V$_G$=-9 V (yellow line) to V$_G$=9 V (blue line). (b) The same data set, after removal of the voltage crosstalk (color coding is the same). In the next step the ordinary Hall effect background has been removed, and the obtained Hall slope R$_H$ around zero external magnetic field for all of the investigated curves is presented in (c) (blue points). As a consistency test the expected slope value calculated from the fitting parameters obtained from the high field analysis in Fig.S8 is also plotted in (c) (red points). (d)(e)(f) Hall magnetic field sweeps after removal of the ordinary Hall effect. (g)(h) Raw longitudinal magneto-resistivity corresponding to the Hall dataset in (a). (i) and (j) show color maps of the Hall resistance magnetic field sweeps from $\mu_{0}H=1~T$ to $\mu_{0}H=-1~T$ for a full range of investigated gate voltages, respectively before (i) and after (j) the ordinary Hall slope removal.
}%
\label{fig:Figs6}%
\end{figure}

\begin{figure}[h]%
\includegraphics[width=\columnwidth]{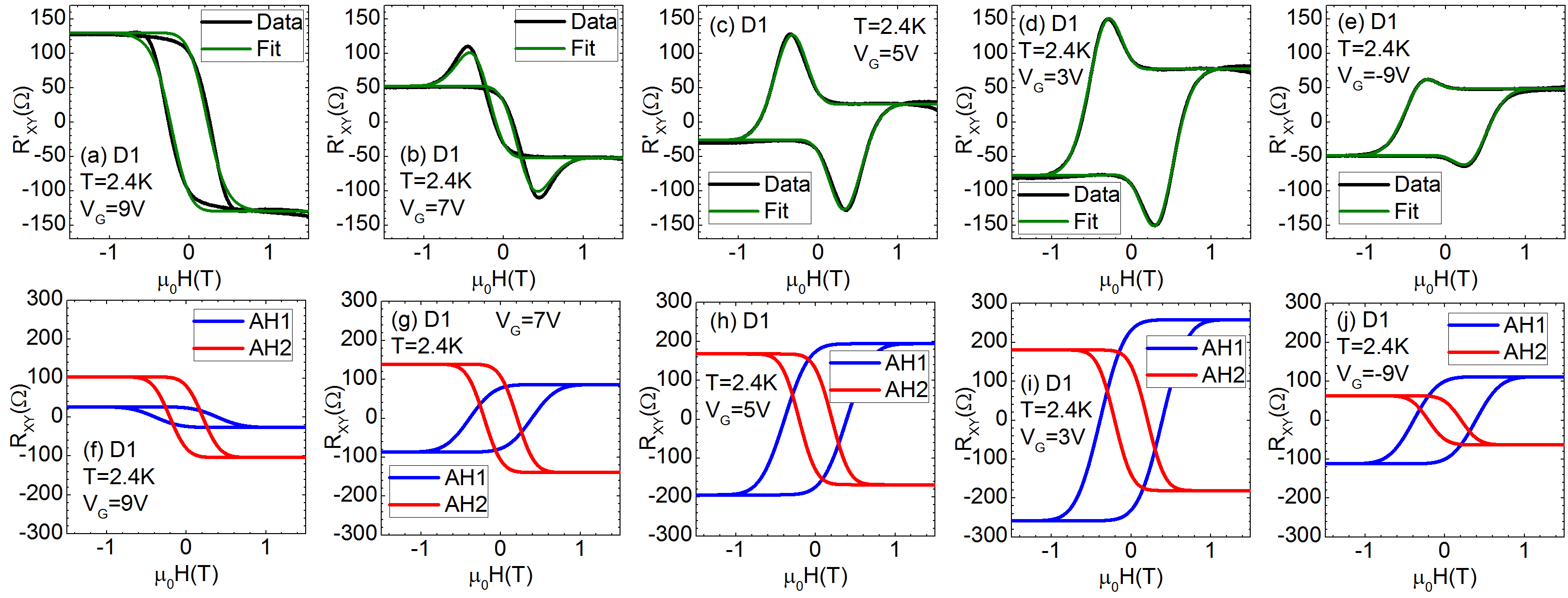}
\caption{
Hall resistance magnetic field sweeps R'$_{xy}$ for different fixed gate voltage values, collected from device D1 at 2.4 K are plotted in (a) (V$_G$=9 V), (b) (V$_G$=7 V), (c) (V$_G$=5 V), (d) (V$_G$=3 V), and (e) (V$_G$=-9 V). The black lines represents experimental data, and olive lines are simulated curves using the two component model. The two anomalous Hall contributions obtained from the fitting are plotted separately for clarity below in (f) (V$_G$=9 V), (g) (V$_G$=7 V), (h) (V$_G$=5 V), (i) (V$_G$=3 V), and (j) (V$_G$=-9 V), where the red color depicts the negative contribution, and the blue color shows the positive contribution. 
}%
\label{fig:Figs7}%
\end{figure}

\begin{figure}[h]%
\includegraphics[width=\columnwidth]{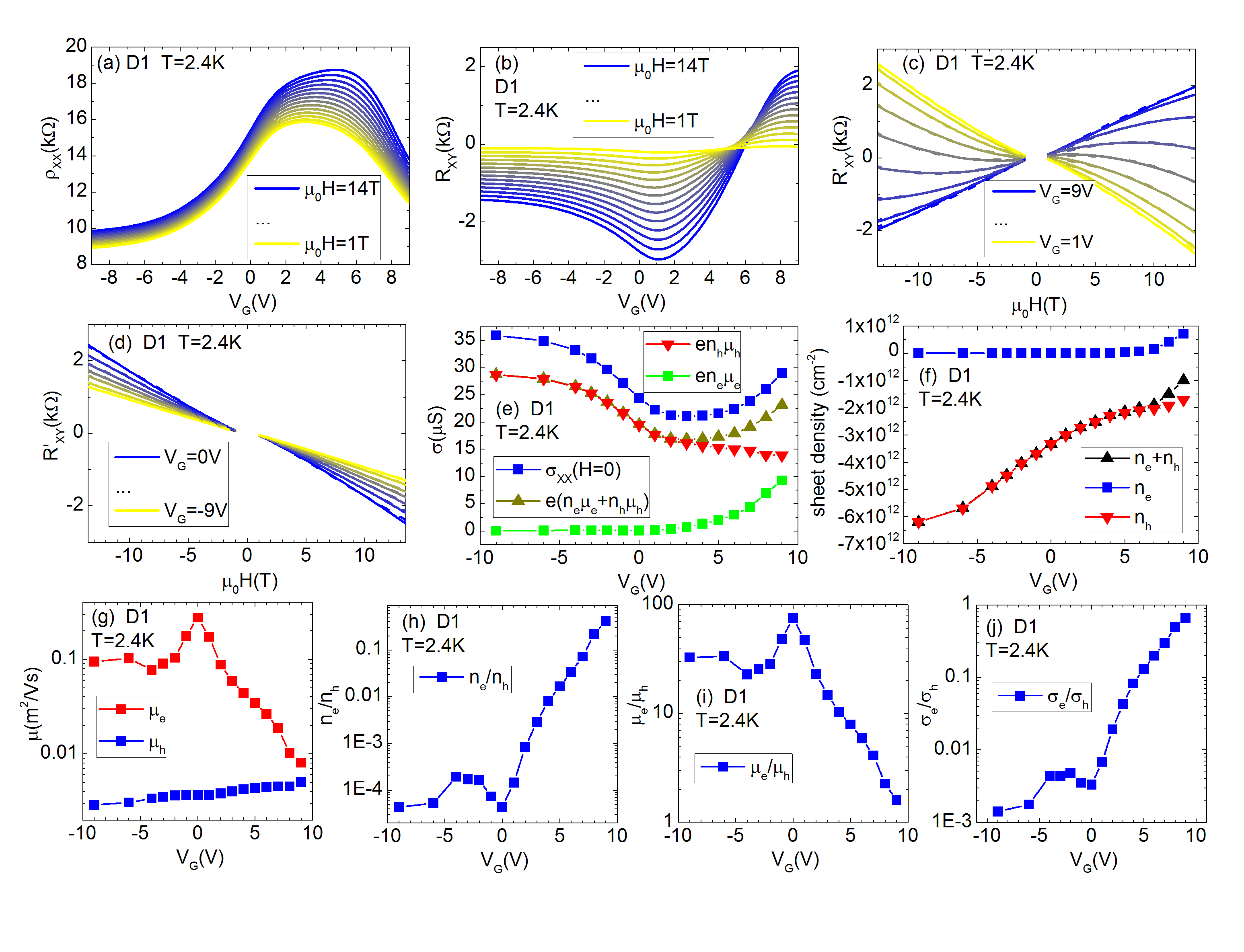}
\caption{
High magnetic field analysis of the device D1 collected at 2.4 K. Raw $\rho_{xx}$ (a) and R$_{xy}$ (b) gate voltage sweeps collected for fixed external magnetic field values ranging from $\mu_{0}$H=14 T to $\mu_{0}$H=1 T in intervals of 1 T. R'$_{xy}$ curves after the voltage crosstalk correction were plotted as a function of external magnetic field in (c) and (d). (e) Drude conductivity of holes (red points), electrons (green points), as well as sum of electrons and holes conductivity (dark yellow). (f) Black points show the total carrier density, blue and red points show the obtained sheet densities respectively for electrons and holes. (g) Mobilities of electrons (red points) and holes (blue points). (h) Ratio of electron and hole sheet density. (i) Ratio of electron and hole mobilities. (j) Ratio of electron and hole Drude conductivity. 
}%
\label{fig:Figs8}%
\end{figure}

\begin{figure}[h]%
\includegraphics[width=\columnwidth]{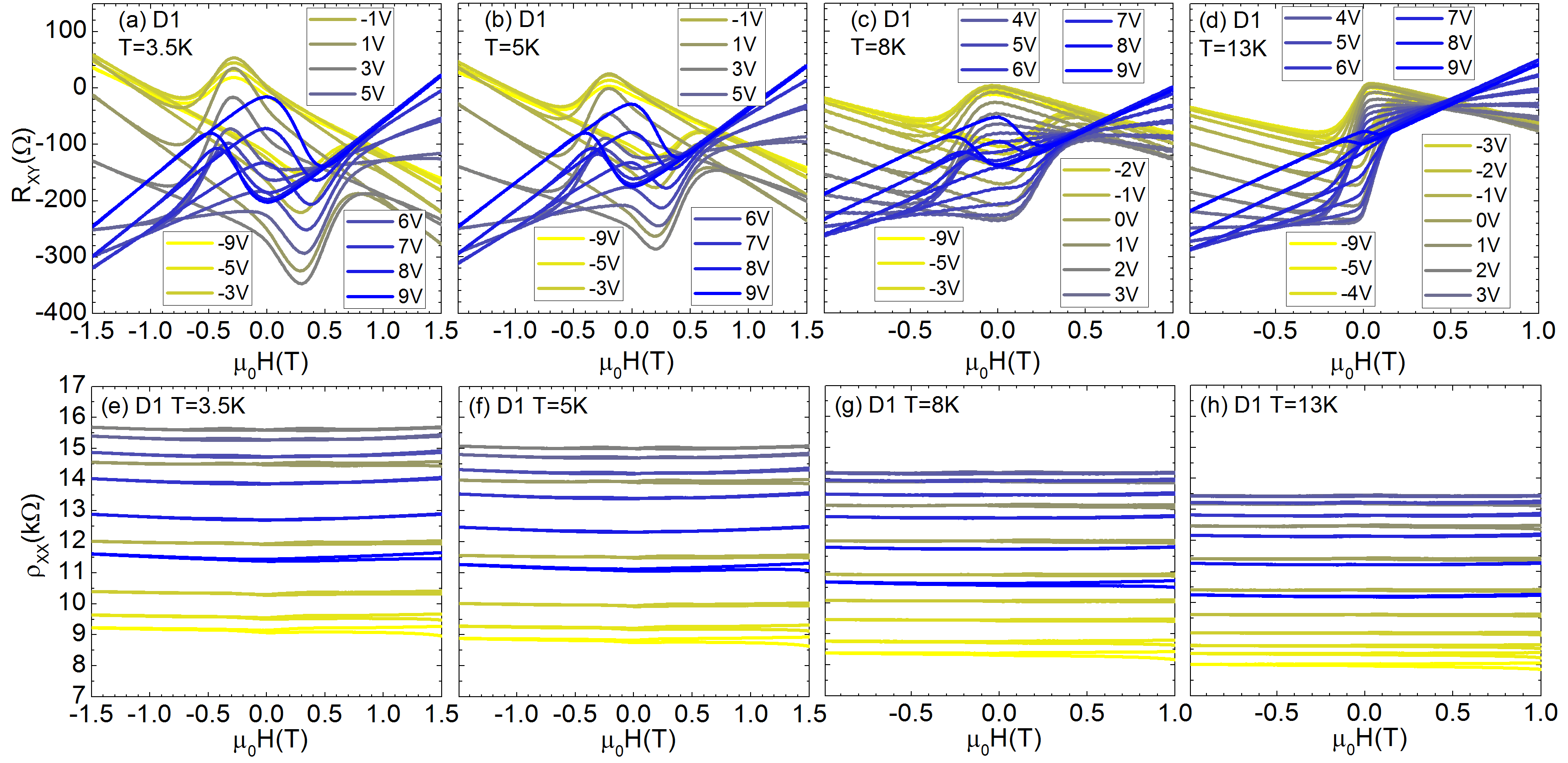}
\caption{
Raw experimental data from device D1 collected at higher fixed temperature values. Hall magneto-resistance collected at the temperature 3.5 K (a), 5 K (b), 8 K (c), and 13 K (d), for a full range of fixed gate voltage values. Below are plotted respective $\rho_{xx}$ data collected at the temperature 3.5 K (e), 5 K (f), 8 K (g), and 13 K (h).
}%
\label{fig:Figs10}%
\end{figure}

\begin{figure}[h]%
\includegraphics[width=\columnwidth]{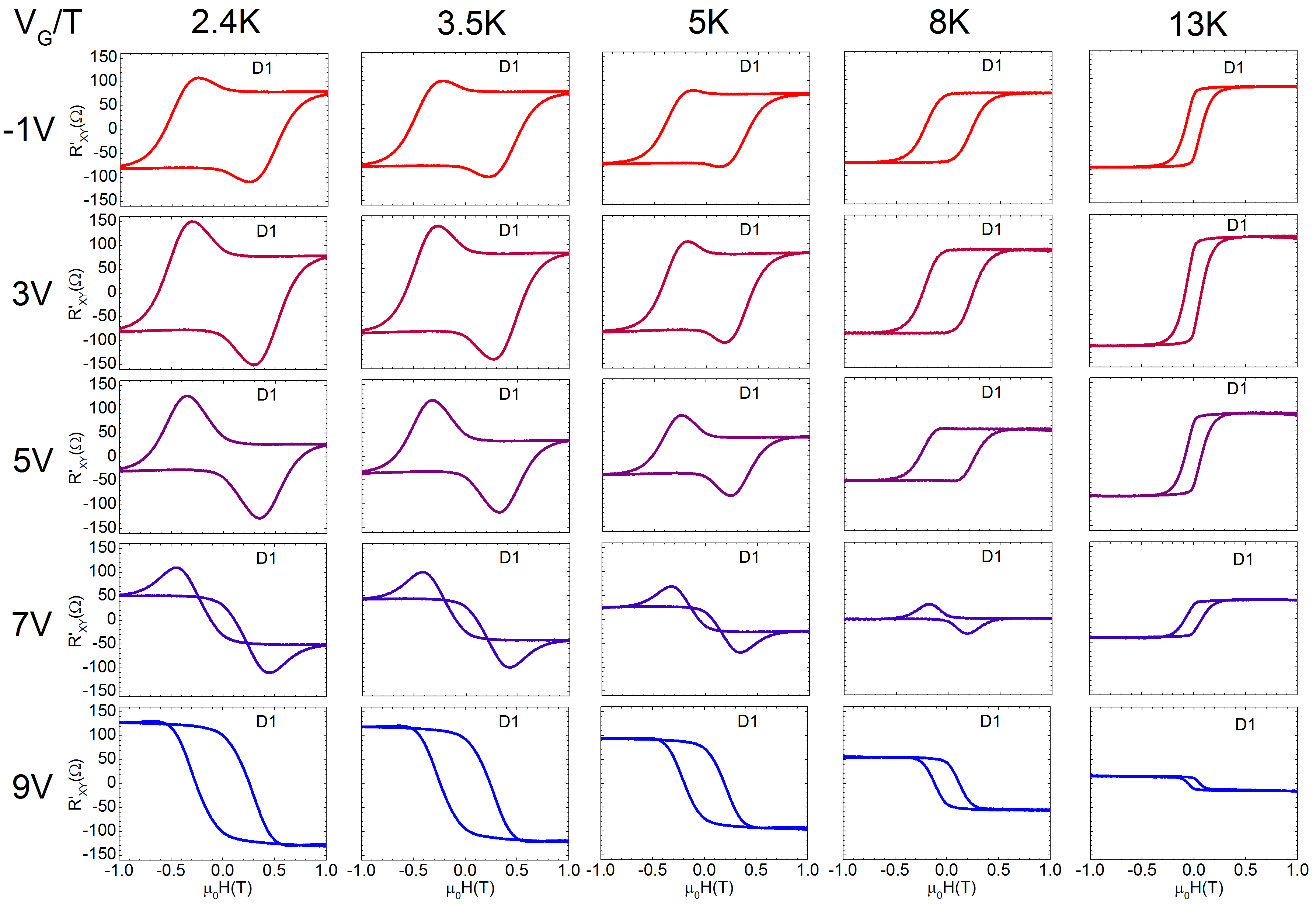}
\caption{
R'$_{xy}$ evolution with the gate voltage and temperature for device D1. The horizontal rows are 2.4 K, 3.5 K, 5 K, 8 K, and 13 K. The range and values for both axis are the same in all of the figures.
}%
\label{fig:Figs11}%
\end{figure}

\end{document}